\documentclass[a4paper,twocolumn,showpacs,preprintnumbers,aps,amsmath,amssymb,nofootinbib]{revtex4}
\input psfig.sty 
\usepackage{picinpar}
\usepackage[dvips]{graphicx}

\begin{document}

\title{An interacting model for the cosmological dark sector}

\author{F. E. M. Costa$^{1,2}$\footnote{E-mail: ernandes@on.br}}

\author{J. S. Alcaniz$^{1}$\footnote{E-mail: alcaniz@on.br}}

\author{Deepak Jain$^{3}$\footnote{E-mail: djain@ddu.du.ac.in}}

\affiliation{$^{1}$Observatorio Nacional, 20921-400, Rio de Janeiro - RJ, Brazil}

\affiliation{$^{2}$Departamento de Astronomia, Universidade de S\~ao Paulo, 05508-900, S\~ao Paulo, SP, Brazil}

\affiliation{$^{3}$Deen Dayal Upadhyaya College, University of Delhi, New   Delhi 110 015, India}

\date{\today}

\begin{abstract}

We discuss a new interacting model for the cosmological dark sector in which the attenuated dilution of cold dark matter scales as $a^{-3}f(a)$, where $f(a)$ is an arbitrary function of the cosmic scale factor $a$. From thermodynamic arguments, we show that $f(a)$ is proportional to entropy source of the particle creation process. In order to investigate the cosmological consequences of this kind of interacting models, we expand $f(a)$  in a power series and viable cosmological solutions are obtained. Finally, we use current observational data to place constraints on the interacting function $f(a)$.

\end{abstract}

\maketitle

\section{Introduction}

A better understanding of the physical mechanism behind the current cosmic acceleration is one of the major challenges both for the cosmology and fundamental physics.
In the framework of the general theory of relativity, the cosmic acceleration can be explained either if a new hypothetical energy component with negative pressure
(dark energy) dominates the current composition of the cosmos~\cite{review} or if the matter content of the universe is subject to dissipative processes (see, e.g. \cite{dissip}).

Following the first route, the simplest and most natural candidate for dark energy is the cosmological constant, $\Lambda$, which corresponds to the energy density stored
in the true vacuum state of all existing fields in the Universe. From the observational point of view, it is well known that the $\Lambda$CDM model is in good agreement
with almost all sets of cosmological observations. Despite its observational successes, it suffers at least from two problems. First, and possibly the most serious one
is the cosmological constant problem (CCP) in which the cosmological upper bound ($\rho_{\Lambda} \lesssim 10^{-47}$ ${\rm{GeV}}^4$) differs from theoretical expectations
($\rho_{\Lambda} \sim 10^{71}$ ${\rm{GeV}}^4$) by more than 100 orders of magnitude. The other is known as coincidence problem and consists to understand why $\rho_{\Lambda}$
is not only small, but also of the same order of magnitude of the energy density of cold dark matter (CDM)~\cite{weinberg}.

An attempt to alleviate the cosmological constant problems is to allow the vacuum energy and dark matter to interact, leading to the so-called vacuum decay models for which
$\Lambda$ is necessarily a time-dependent quantity. Cosmological scenarios with a dynamical $\Lambda$ term were independently proposed about two decades ago in
Refs.~\cite{lista}
(see also~\cite{cq,jesus,ernandes1,costa82} for interacting models in the which the dark energy is represented by a smooth component parametrized by an equation of
state $p_{\rm{DE}} = w \rho_{\rm{DE}}$
with $w < 0$). Most of these scenarios have specific decay laws for $\Lambda$ which lead to particular cosmological solutions and phenomenology (see, e.g., Table I of
\cite{overduin} for some proposed $\Lambda$ decay laws).

In this paper, we discuss a general cosmological scenario of $\Lambda$-CDM interaction in which the attenuated dilution of CDM is described by an arbitrary function of
the cosmic scale factor $f(a)$. We show that this class of models has many of the previous phenomenological scenarios as a particular case. Using thermodynamic considerations,
it is shown that the function $f(a)$ is proportional to the entropy source of the associated particle creation process. We discuss some interesting cosmological consequences of
this $\Lambda$-CDM interaction, such as a possible loitering phase in a spatially flat universe and carry out a joint statistical analysis with recent observations of type Ia
supernovae, baryonic acoustic oscillations and cosmic microwave background data to test its observational viability. We work in units where $(8\pi G)^{1/2} = c = 1$.

\section{$\Lambda$-dark matter interaction}

According to the Bianchi identities, the Einstein field  equations, ${G}^{\mu \nu} = {T}^{\mu \nu} + \Lambda {g}^{\mu \nu}$, implies that $\Lambda$ is necessarily a
constant either if the energy-momentum tensor of matter fields and CDM particles $T^{\mu \nu}$ is null or if it is separately conserved, i.e.,
$u_\mu {{T}}^{\mu \nu};_{\nu} =  0$. This amounts to saying that the presence of a time-varying cosmological term results
in a coupling between $\Lambda$ and other cosmic component. By considering the Friedmann-Lemaitre-Robertson-Walker space-time and a coupling between $\Lambda$ and CDM particles,
we have that
\begin{equation} \label{ec}
\dot{\rho}_{dm} + 3\frac{\dot{a}}{a}\rho_{dm} = - \dot{\rho}_\Lambda = -Q\;,
\end{equation}
where $\rho_{dm}$ and $\rho_\Lambda$ are the energy densities of CDM and $\Lambda$ respectively and the dot sign denotes derivative with respect to the time.

As the cosmological term interacts with CDM particles, their energy density will dilute slowly compared with the standard relation, $\rho_{dm} \propto a^{-3}$. Thus,
the deviation from the standard dilution may be characterized by the function $f(a)$, such that
\begin{equation}\label{ansatz}
\rho_{dm}=a^{-3}f(a).
\end{equation}
By substituting the relation (\ref{ansatz}) into (\ref{ec}), we find
\begin{equation}\label{dla}
\frac{d\rho_{\Lambda}}{da} = - a^{-3}\frac{df}{da}\;,
\end{equation}
where for $f(a) = \mbox{const.} = \rho_{dm,0}$ the standard $\Lambda$CDM scenario is fully recovered. Note that for $f(a) = \rho_{dm,0}a^{\epsilon}$ the dynamical $\Lambda(t)$CDM scenarios recently discussed in Refs.~\cite{wm,alclim05,ernandes,costa81} are obtained whereas the vacuum decay model with $\Lambda \propto H$ investigated in Refs. \cite{saulo} is recovered when $f(a) = c_1 + c_2 a^{3/2}$. More general coupled quintessence scenarios in which the coupling is written in terms of the scalar field $\phi$ as $Q \equiv \sqrt{2/3} \beta \rho_{dm}\dot{\phi}$~\cite{amen04} (see also \cite{petto08}), where $\beta$ is a coupling function, are recovered by identifying
\begin{equation}\label{comp}
\frac{\dot{f}}{f} = -\sqrt{2/3} \beta \dot{\phi}.
\end{equation}

In this paper, we will work with the $f(a)$ function as a power series of scale factor, i.e.,
\begin{equation}\label{expan}
f(a) = f_0 + f_1 a + f_2 a^{2} +...,
\end{equation}
where $f_0$, $f_1$ and $f_2$ are constants. From Eqs. (\ref{ansatz}), (\ref{dla}) and (\ref{expan}), we find
\begin{equation}\label{serie}
\rho_{dm}= f_0 a^{-3} + f_1 a^{-2} + f_2 a^{-1} +...,
\end{equation}
\begin{equation}\label{an}
\rho_{\Lambda}= \tilde{\rho}_{\Lambda,0} + \frac{f_1}{2} a^{-2}  + 2f_2 a^{-1} +...,
\end{equation}
where $\tilde{\rho}_{\Lambda,0}$ is an integration constant.

Now, neglecting the radiation contribution and considering that the baryonic content is separately conserved, the expansion rate of the Universe can be written as
\begin{equation}\label{aga}
{\cal{H}} =  \tilde{\Omega}_{\Lambda,0} + (\Omega_0 + \Omega_{b,0})a^{-3} + \frac{3}{2}\Omega_1  a^{-2} + 3\Omega_2 a^{-1} + ... ,
\end{equation}
where ${\cal{H}} = (H/H_0)^2$, $\tilde{\Omega}_{\Lambda,0}=\tilde{\rho}_{\Lambda,0}/\rho_{c,0}$, $\Omega_{b,0} = \rho_{b,0}/\rho_{c,0}$ represents the density parameter of baryons, $\Omega_0 = f_0 /\rho_{c,0}$, $\Omega_i = f_i /\rho_{c,0}$ (with $i = 1, 2,...,n$) and $\rho_{c,0}= 3H_{0}^{2}$ is the present critical energy density. In what follows, we will restrict our analysis up to the first order expansion of $f(a)$. The practical reason for
this choice is shown in Figure 1 where the ratio $\rho_{\Lambda}/\rho_{dm}$ as a function of the redshift is displayed for first (solid line) and second order expansions
in $f(a)$ assuming the best-fit values of $\Omega_0$, $\Omega_1$ and $\Omega_2$ discussed in Sec. IV. Clearly, higher order terms do not modify the ratio between $\Lambda$
and the CDM component for the  interval considered. We note that an expansion rate similar to Eq. (\ref{aga}) has been explored in reconstruction studies of the dark energy
density (see, e.g., Ref.~\cite{sahni1}).

\section{Thermodynamics of interaction}

\begin{figure}[t]
\centerline{\psfig{figure=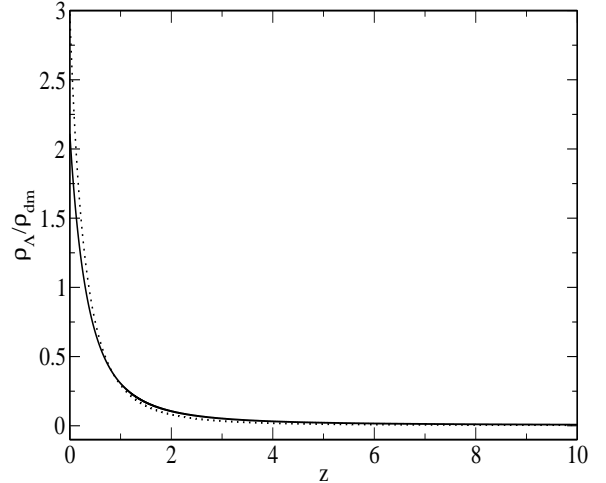,width=3.4truein,height=3.2truein,angle=-90}}
\caption{The ratio of the energy densities as a function of the redshift. It was used the best-fit values of $\Omega_0$, $\Omega_1$ and $\Omega_2$.}
\end{figure}

Let us now consider some thermodynamic aspects of interaction between the vacuum and CDM particles. The thermodynamic behavior of a decaying vacuum system is simplified
either if one assumes that the chemical potential of the vacuum component is zero or if the vacuum medium plays the role of a condensate carrying no entropy, as happens
in the two fluid description employed in the super fluid thermodynamics \cite{lima}.

In this case, the thermodynamic description requires only the knowledge of the particle flux, $N^{\alpha} = n u^{\alpha}$, and entropy flux, $S^{\alpha} = n\sigma u^{\alpha}$,
where $n=N/a^{3}$ and $\sigma = S/N$ are, respectively, the concentration and the specific entropy (per particle) of the created component. The equation for $n$ has necessarily
a source term, i.e.,
\begin{equation}\label{conc}
\dot{n} + 3\frac{\dot{a}}{a} n = \Psi = n\Gamma
\end{equation}
where $\Psi$ is particle source $(\Psi > 0)$, or a sink $(\Psi < 0)$, and $\Gamma$ is the rate of change of particle number. Since $\rho_{dm} = nm$, Eq. (\ref{ansatz}) gives
\begin{equation}\label{con}
n=n_{0}{a^{-3}}f(a)\;.
\end{equation}
Substituting above equation in Eq. (\ref{conc}), it follows that
\begin{equation}\label{taxa}
\dot{f} = f\Gamma.
\end{equation}
As argued in Ref. \cite{lima}, the energy exchange between the vacuum and the cold dark matter may take place in several ways. The most physically relevant case has been
termed adiabatic decaying
vacuum. In this case, several equilibrium relations are preserved, and perhaps  more importantly the entropy of created particles increases wheras the specific entropy
(per particle) remains constant
$(\dot{\sigma}=0)$. This mean that
\begin{equation}\label{entro}
\frac{\dot{S}}{S} = \frac{\dot{N}}{N} = \Gamma.
\end{equation}
From Eqs. (\ref{taxa}) and (\ref{entro}) it is straightforward to show that
\begin{equation}
f \propto S.
\end{equation}
The second law of thermodynamics, therefore, implies that $f(a)$ must be an increasing function of the scale factor, which requires $\Omega_1 > 0$. Physically, this
result indicates that only an energy flow from vacuum to cold dark matter is allowed. As discussed in the next section,
this is in disagreement with current observational constraints that favor an energy flow from dark matter to vacuum $(\Omega_1 < 0)$.

\section{Observational analysis}

\begin{figure}[t]
\centerline{\psfig{figure=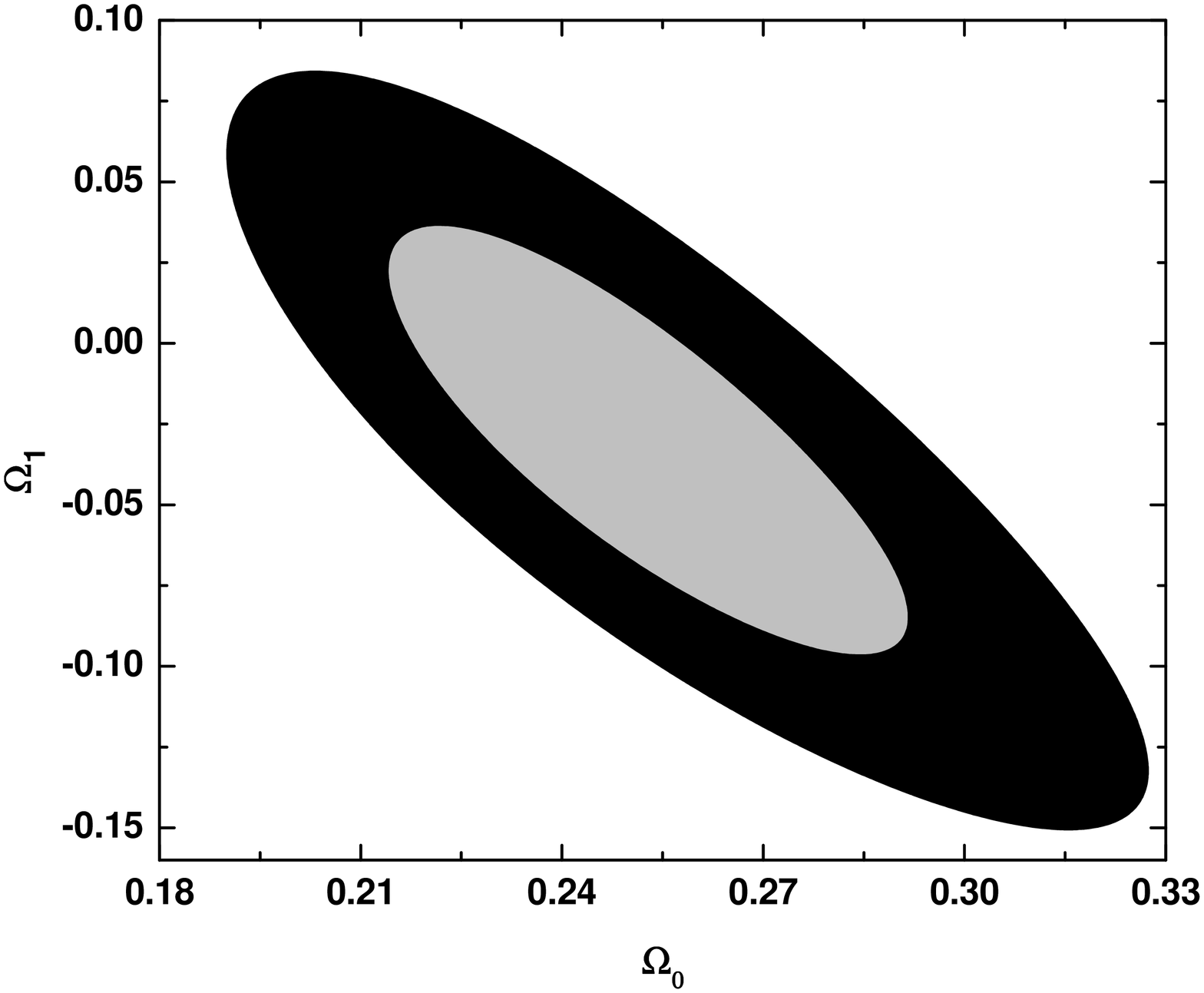,width=3.2truein,height=3.4truein,angle=-90}}
\caption{The results of our statistical analysis. Contours of $\chi^2$ in the plane $\Omega_0 - \Omega_1$. These contours are drawn for $\Delta \chi^2 = 2.30$ and $6.17$.}
\end{figure}

In order to delimit the parametric space $\Omega_0 - \Omega_1$ we perform a joint analysis involving current SNe Ia, BAO and CMB data. In our analysis, we fix
$\Omega_{b,0} = 0.0416$ from WMAP results~\cite{cmbnew} (which is also in good agreement with the bounds on the baryonic component derived from primordial
nucleosynthesis~\cite{nucleo}) and marginalize over the Hubble parameter $H_0$.

We use one of the most recent SNe Ia compilation, the so-called Union 2 sample compiled in Ref.~\cite{union} which includes 557 data points after selection cuts.
Following Ref.~\cite{sollerman} we also use measurements derived from the product of the CMB acoustic scale $\ell_{A} = \pi d_A (z_*)/r_s(z_*)$ and from the ratio of
the sound horizon scale at the drag epoch to the BAO dilation scale, $r_s(z_d)/D_V(z_{\rm{BAO}})$, where $d_A (z_*)$ is the comoving angular-diameter distance to
recombination ($z_* = 1089$) and $r_s(z_*)$ is the comoving sound horizon at photon decoupling.
In the above expressions,  $z_d \simeq 1020$ is the redshift of the drag epoch (at which the acoustic oscillations are frozen in) and the dilation scale, $D_V$, is
given by $D_V(z) = [zr^{2}(z)/H(z)]^{1/3}$. By combining the ratio $r_s (z_d = 1020)/r_s (z_*=1090) = 1.044 \pm 0.019$ ~\cite{Komatsu,Percival} with the measurements of $r_s(z_d )/D_V(z_{\rm{BAO}})$ at $z_{\rm{BAO}} =$ 0.20, 0.35 and 0.6.,
one finds
$
f_{0.20} = d_A (z_*)/D_V (0.2) = 18.32 \pm0.59,
$
$
f_{0.35} = d_A (z_*)/D_V (0.35) = 10.55 \pm 0.35\;,
$
$
f_{0.6} = d_A (z_*)/D_V (0.60) = 6.65 \pm 0.32\;
$~\cite{sollerman,blake}.

In our analysis, we minimize the function $\chi^2_{\rm{T}} = \chi^2_{\rm{SNe}} + \chi^2_{\rm{CMB/BAO}}$, where $\chi^2_{\rm{SNe}}$ and $\chi^2_{\rm{CMB/BAO}}$ correspond
to the SNe Ia and CMB/BAO $\chi^2$ functions, respectively. Figure 2 shows confidence regions (68.3\% CL and 95.4\%) in $\Omega_0 - \Omega_1$ plane obtained from this joint
analysis. At $2\sigma$ we found $\Omega_{0} = 0.25_{-0.03}^{+0.03}$ and $\Omega_{1}= - 0.028_{-0.05}^{+0.05}$. As expected, we found values of $\Omega_{1} \simeq 0$
since it acts as a curvature term ($a^{-2}$) in Eq. (\ref{aga}). As shown
in Section III, the interacting parameter $\Omega_{1}$ is restricted from thermodynamical arguments to be positive. By considering this physical constraint on $\Omega_{1}$ and minimizing the function $\chi^2_{\rm{T}}$, we obtain $\Omega_{0} = 0.24_{-0.03}^{+0.03}$ and $\Omega_{1} = 0.0_{-0.00}^{+0.04}$ at $2 \sigma$ level. An observational comparison between the model here described and the coupled quintessence scenario of  Ref.~\cite{amen04} can be made by comparing the above estimates for $\Omega_{1}$ with the constraints on the coupling function $\beta$ [see Eq. (\ref{comp})]. For the case in which $\beta$ remains constant with $z$, Ref.~\cite{amend00} used LSS and CMB data to found $|\beta| < 0.1$, which is in full agreement with the bounds on $\Omega_{1}$ obtained in our analysis.

\begin{figure*}
\centerline{\psfig{figure=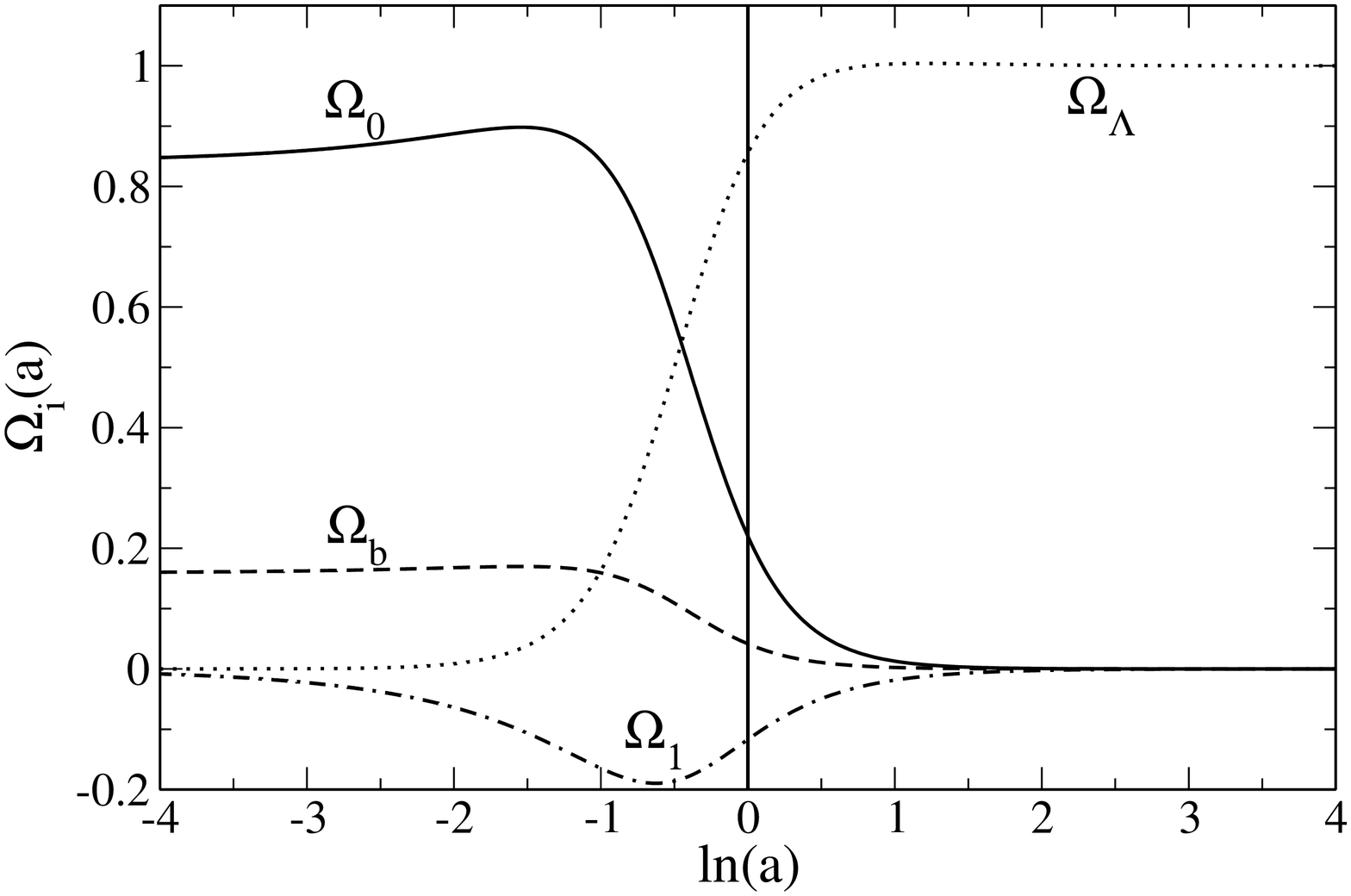,width=2.4truein,height=2.2truein,angle=-90}
\psfig{figure=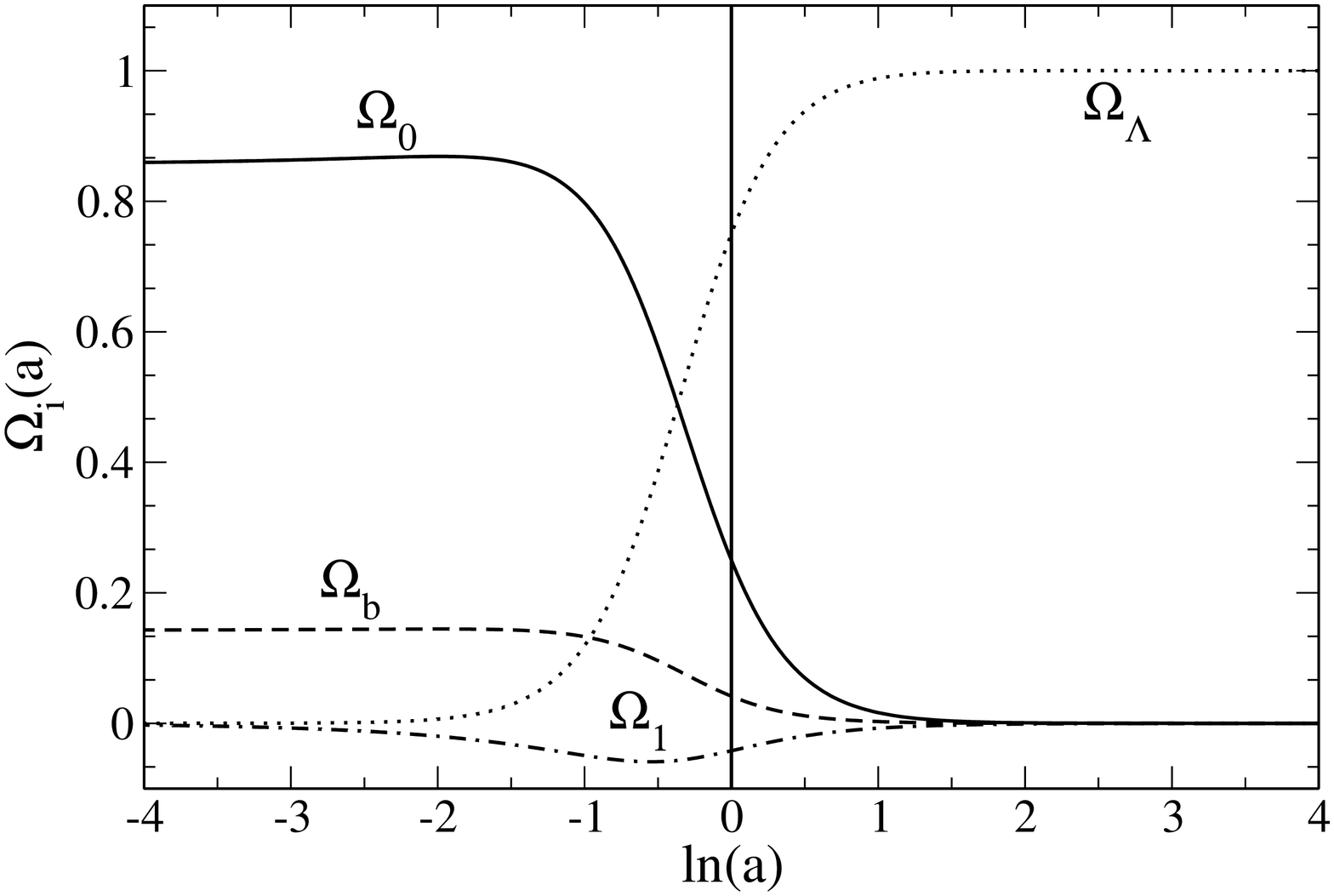,width=2.4truein,height=2.2truein,angle=-90}
\psfig{figure=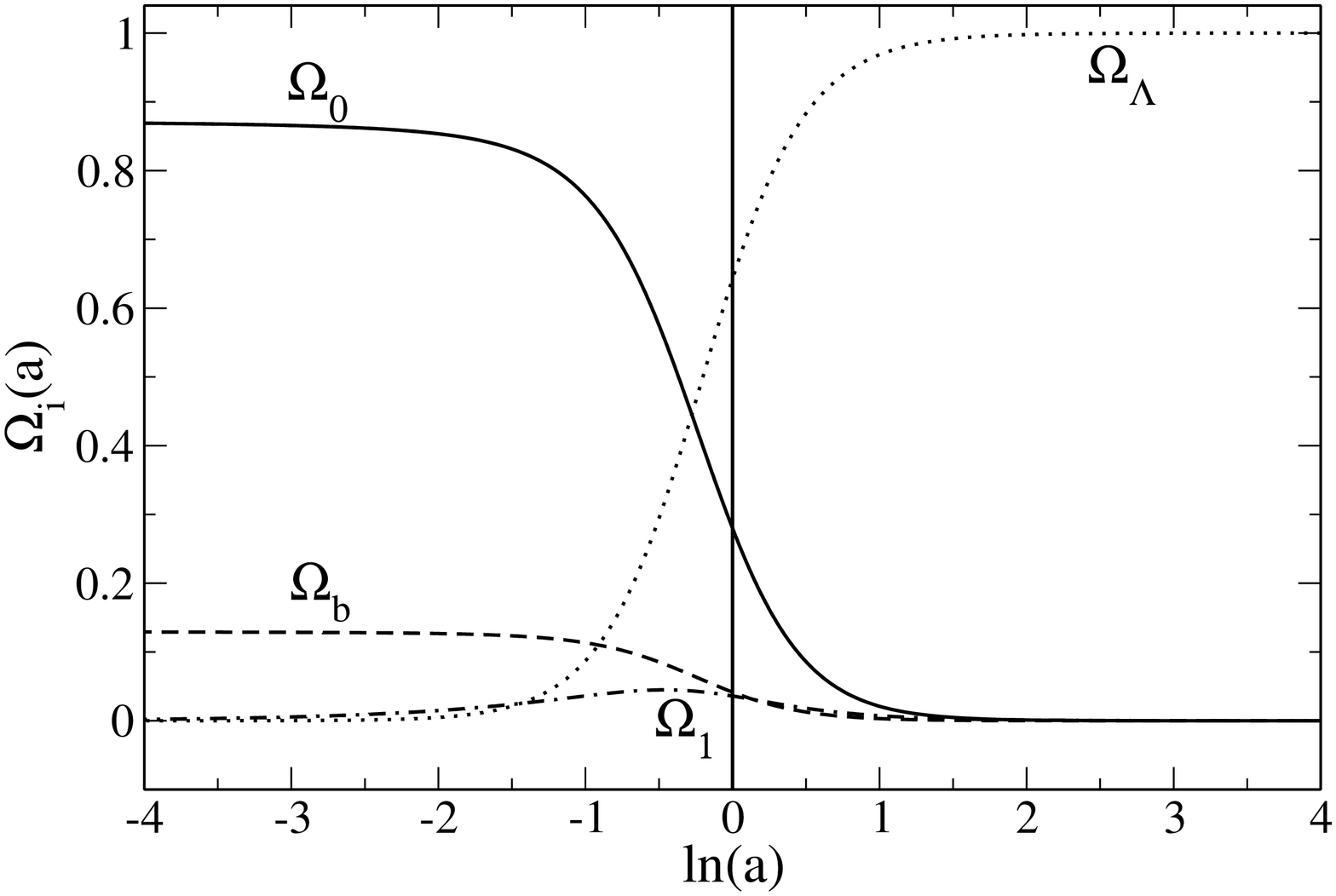,width=2.4truein,height=2.2truein,angle=-90}}
\caption{Evolution of the density parameters  as a function of $\ln(a)$ for the lower limit (left panel), best-fit (central panel) and upper limit (right panel) on
$\Omega_0$ and $\Omega_1$ obtained from the statistical analysis described above.}
\label{fig:qzw}
\end{figure*}

For the sake of completeness, we also derive the time evolution of the density parameters $\Omega_b(a)$, $\Omega_{0}(a)$, $\Omega_{1}(a)$ and $\Omega_{\Lambda}(a)$ which
are given, respectively, as
$$
\Omega_{i}(a) = \frac{\Omega_{i,0} a^{-3}}{\xi(a)},\quad \Omega_{1}(a) = \frac{1.5 \Omega_{1}a^{-2}} {\xi(a)},\quad \Omega_{{\rm{\Lambda}}}(a) =
\frac{\tilde{\Omega}_{\Lambda,0}} {\xi(a)}\;,
$$
where
$$
\xi(a) = {\tilde{\Omega}_{\Lambda,0} + (\Omega_0 + \Omega_{b,0})a^{-3} + \frac{3}{2}\Omega_1  a^{-2}}
$$
and $i = b, 0$. Figure 3 shows the evolution of $\Omega_b(a)$, $\Omega_{0}(a)$, $\Omega_{1}(a)$ and $\Omega_{\Lambda}(a)$ with the logarithm of the scale factor
$\ln(a)$ for $\Omega_{b,0}=0.0416$ and $\Omega_0 = 0.22$, $0.25$, $0.28$ and $\Omega_1 = -0.078$, $-0.028$, $0.022$, which correspond to lower limit, best-fit and upper
limit (at $2\sigma$ of confidence) from statistical analysis described above. In all cases a mix of baryons
($\lesssim 20\%$) and dark matter ($\gtrsim 80\%$) dominates the past evolution of the Universe whereas the contributions relative to the integration constant
$\tilde{\Omega}_{\Lambda,0}$ and the interaction parameter $\Omega_1$ vanish at high-z. As expected, and in agreement with observational data, we note that
$\tilde{\Omega}_{\Lambda,0}$ is the dominant component from a value of $a_* \lesssim 1$ on, while the interaction parameter $\Omega_1 (a)$ is always
the subdominant throughout cosmic evolution.

Another interesting aspect of this flat vacuum decay model is that it can accommodate a loitering phase. As discussed in Ref. \cite{sahni92}, a loitering phase is
characterized by a $dH/dz = 0$. Deriving Eq. (\ref{aga}) and using the loitering condition, we find ($1+z=a^{-1}$)
\begin{equation}%
1+ z_l = \frac{|\Omega_1 |}{\Omega_0 + \Omega_{b,0}}.
\end{equation}
In the context of this flat interacting model, the existence of a recent loitering phase is possible only if $|\Omega_1 | > \Omega_0 + \Omega_{b,0}$. By considering the
obsevational limits on these parameters given above, we note that a realistic loitering phase is possible only in the future ($z_l < 0$). Finally, by considering that
$\ddot{a}/{a} = \dot{H} + H^{2}$, it follows that $(\ddot{a}/{a})_{z = z_l} = H^{2}(z = z_l )$, which clearly shows that models with a loitering phase are
compatible with an accelerating universe, as indicated by SNe Ia data.\\

\section{Conclusions}

We have discussed cosmological consequences of a new cosmological scenario of vacuum decay in which the deviation of standard evolution of the cold dark matter
component is characterized by an arbitrary function of the cosmic scale factor $f(a)$. Using thermodynamic arguments, we have shown that the interaction function
$f(a)$ is proportional to the entropy source of the particle creation process, which means that only an energy flow from vacuum to cold dark matter is allowed.  We have
investigated the observational viability of this class of models from recent data of SNe Ia and the distance ratio from baryon acoustic oscillation at different redshifts
and CMB decoupling at $z_{\rm{LS}} = 1089$. In particular, when  $f(a)$ can be expanded in a power series of the scale factor, we find strong constraints on the parameters
of this model, with a slight  deviation from standard $\Lambda$CDM dynamics.

Finally, it is worth emphasizing that we have restricted the present analysis to coupled quintessence models in which $w_\phi = -1$ (dynamical $\Lambda$ models,
whereas a full treatment of the dark matter-dark energy interaction must also take into account the role of the dark energy equation-of-state in the process. Some theoretical
and observational consequences of a $w$-CDM interacting scenario with a time-dependent coupling term, as well as a scalar field description for this class of models is currently under investigation.

\begin{acknowledgments}

FEMC acknowledges financial support from CNPq and FAPESP (2011/13018-0). DJ acknowledges the financial support provided by Department of Science and Technology (DST), Govt. of India, under the project No. SR/S2/HEP-002/008. JSA acknowledges financial support from CNPq (305857/2010-0) and FAPERJ (E-26/103.239/2011).

\end{acknowledgments}


\begin{thebibliography}{}

\bibitem{review} V.~Sahni and A.~A.~Starobinsky, Int.\ J.\ Mod.\   Phys.\  D {\bf 9}, 373 (2000); P. J. E. Peebles and B. Ratra  Rev. Mod. Phys. {\bf{75}}, 559 (2003);  T. Padmanabhan,  Phys. Rept. {\bf{380}}, 235 (2003); E. J. Copeland, M. Sami and   S. Tsujikawa, Int. J. Mod. Phys. {\bf{D15}}, 1753 (2006);   J. S. Alcaniz, Braz. J. Phys. {\bf{36}}, 1109 (2006). [astro-ph/0608631]; J.~A.~Frieman,  AIP Conf.\ Proc.\  {\bf 1057}, 87 (2008). arXiv:0904.1832 [astro-ph.CO].

\bibitem{dissip}  J.~A.~S.~Lima and J.~S.~Alcaniz,  Astron.\ Astrophys.\  {\bf 348}, 1 (1999);  J.~S.~Alcaniz and J.~A.~S.~Lima,  Astron.\ Astrophys.\  {\bf 349}, 729 (1999); L. P. Chimento, A. S. Jakubi, D. Pavon, and W. Zimdahl, Phys. Rev. D67, 083513 (2003).

\bibitem{weinberg} S.~Weinberg, arXiv:0005265 [astro-ph.CO].

\bibitem{lista}  M. $\ddot{\rm{O}}$zer and M. O. Taha, Phys. Lett. B   {\bf 171}, 363 (1986); Nucl. Phys. B {\bf 287}, 776 (1987);  O. Bertolami, Nuovo Cimento Soc. Ital. Fis., {\bf{B93}}, 36 (1986);  K. Freese {\it et al.}, Nucl. Phys. {\bf{B287}}, 797 (1987); W. Chen  and Y-S. Wu, Phys. Rev. D {\bf 41}, 695 (1990); D. Pav\'{o}n,  Phys. Rev. D {\bf 43}, 375 (1991); J. C. Carvalho, J. A. S. Lima and  I. Waga, Phys. Rev. D {\bf{46}}, 2404 (1992);  J. S. Alcaniz and  J. M. F. Maia, Phys. Rev. {\bf{D67}}, 043502 (2003); E. Elizalde,  S. Nojiri, S.D. Odintsov and P. Wang, Phys. Rev. D {\bf{71}}, 103504  (2005); J.~Grande, J.~Sola and H.~Stefancic,  JCAP {\bf 0608}, 011  (2006).

\bibitem{cq} D. Wands, E. S. Copeland and A. Liddle,  Ann. N. Y. Acad. Sci. \textbf{688}, 647 (1993);  W. Zimdahl, D. Pav\'{o}n and  L. Chimento,Phys. Lett. B \textbf{521}, 133 (2001);  L. P. Chimento, A. S. Jakubi, D. Pav\'{o}n and W. Zimdahl, Phys. Rev. D \textbf{67}, 083513 (2003); L. Amendola and D. Tocchini-Valentini, Phys. Rev. D \textbf{64}, 043509 (2001); M. Gasperini, F. Piazza and G. Veneziano, Phys. Rev. D \textbf{65}, 023508 (2002); W. Zimdahl and D. Pav\'{o}n, Gen. Rel. Grav. \textbf{35}, 413 (2003); J.~D.~Barrow and T.~Clifton, Phys.\ Rev.\  D {\bf 73}, 103520 (2006); Z.~K.~Guo, N.~Ohta and S.~Tsujikawa, Phys.\ Rev.\  D {\bf 76} (2007) 023508; O.~Bertolami, F.~Gil Pedro and M.~Le Delliou,  Phys.\ Lett.\  B {\bf   654}, 165 (2007);  Q.~Wu, Y.~Gong, A.~Wang and J.~S.~Alcaniz,  Phys.\ Lett.\  B {\bf 659}, 34 (2008).

\bibitem{jesus} J.~F.~Jesus et al.,  Phys.\ Rev.\  D {\bf 78}, 063514 (2008).

\bibitem{ernandes1}  F.~E.~M.~Costa, E.~M.~Barboza and J.~S.~Alcaniz,  Phys.\ Rev.\  D {\bf 79}, 127302 (2009).

\bibitem{costa82} F.~E.~M.~Costa, Phys.\ Rev.\  D {\bf 82}, 103527 (2010).

\bibitem{overduin} J.~M.~Overduin and F~I.~Cooperstock, Phys.\ Rev.\  D {\bf 58}, 043506 (1998).

\bibitem{wm} P. Wang and X. Meng, Class. Quant. Grav. {\bf 22}, 283 (2005).

\bibitem{alclim05} J. S. Alcaniz and J. A. S. Lima, Phys. Rev. D {\bf  72}, 063516 (2005).

\bibitem{ernandes} F.~E.~M.~Costa, J.~S.~Alcaniz and J.~M.~F.~Maia,  Phys.\ Rev.\  D {\bf{77}}, 083516 (2008).

\bibitem{costa81} F.~E.~M.~Costa and J.~S.~Alcaniz, Phys.\ Rev.\  {\bf  D 81}, 043506 (2010).

\bibitem{saulo} H. A. Borges and S. Carneiro, Gen. Relativ. Gravit. 37, 1385 (2005);   S.~Carneiro {\it et al.}, Phys.\ Rev.\  D {\bf 77}, 083504 (2008); C.~Pigozzo {\it et al.},  JCAP {\bf 1108}, 022 (2011).

\bibitem{amen04} L. Amendola, Phys. Rev. {\bf{D 69}}, 103524 (2004).

\bibitem{petto08} V. Pettorino and C. Baccigalupi, Phys. Rev. {\bf{D 77}}, 103003 (2008).

\bibitem{sahni1} U. Alam, V. Sahni and A. A. Starobinsky, JCAP 0702:011, (2007).

\bibitem{lima}  J.~A.~S.Lima, Phys.\ Rev.\  D {\bf 54}, 2571 (1996).

\bibitem{cmbnew} E.~Komatsu {\it et al.}, Astrophys.\ J.\ Suppl.\ {\bf 180}, 330 (2009).

\bibitem{nucleo} G. Steigman, Ann. Rev. Nucl. Part. Sci, {\bf 57}, 463 (2007).

\bibitem{union} R. Amanullah {\it et al.}, Astrophys. J. {\bf 716}, 712 (2010).

\bibitem{sollerman} J. Sollerman {\it et al.}, Astrophys. J. {\bf{703}}, 1374-1385 (2009).

\bibitem{Komatsu} E. Komatsu {\it et al.}, ApJS {\bf{180}}, 330 (2009).

\bibitem{Percival} W. J. Percival {\it et al.}, Mon. Not. Roy. Astron. Soc. {\bf{401}}, 2148 (2010).

\bibitem{blake} C. Blake {\it et al.}, Mon. Not. Roy. Astron. Soc. {\bf{415}}, 2892 (2011).

\bibitem{amend00} L. Amendola,  Phys. Rev. {\bf{D 62}} 043511 (2000).

\bibitem{sahni92}  V. Sahni {\it et al.}, Astrophys. J., 385 (1992); V. Sahni and Y. Shtanov, Phys.\ Rev.\  {\bf D 71}, 084018 (2005).


\end{thebibliography}
\end{document}